\documentstyle[11pt,aaspp4,epsfig]{article}
\slugcomment{To appear in Astrophysical Journal Letters}

\begin{document}

\title{Phase Lag and Coherence Function of X-ray emission from Black 
Hole Candidate XTE J1550-564}
\author{Wei Cui\altaffilmark{1}, Shuang Nan Zhang\altaffilmark{2}, and 
Wan Chen\altaffilmark{3,4}}

\altaffiltext{1}{Center for Space Research, Massachusetts
Institute of Technology, Cambridge, MA 02139; cui@space.mit.edu}

\altaffiltext{2}{Department of Physics, University of Alabama
in Huntsville, Huntsville, AL 35899; zhangsn@email.uah.edu}

\altaffiltext{3}{NASA/Goddard Space Flight Center, Code 661,
Greenbelt, MD 20771; chen@milkyway.gsfc.nasa.gov}

\altaffiltext{4}{also Department of Astronomy, University of Maryland,
College Park, MD 20742}

\begin{abstract}
We report the results from measuring the phase lag and coherence 
function of X-ray emission from black hole candidate (BHC) XTE
J1550-564. These X-ray temporal properties have been recognized to 
be increasingly important in providing important diagnostics of the
dynamics of accretion flows around black holes. For XTE J1550-564, 
we found significant hard lag --- the X-ray variability in high 
energy bands {\em lags} behind that in low energy bands --- associated 
both with broad-band variability and quasi-periodic oscillation
(QPO). However, the situation is more complicated for the QPO: while 
hard lag was measured for the first harmonic of the signal, the 
fundamental component showed significant {\em soft} lag. Such behavior 
is remarkably similar to what was observed of microquasar GRS 1915+105. 
The phase lag evolved during the initial rising phase of the 1998 
outburst. The magnitude of both the soft and hard lags of the QPO 
increases with X-ray flux, while the Fourier spectrum of the
broad-band lag varies significantly in shape. The coherence function 
is relatively high and roughly constant at low frequencies, and begins 
to drop almost right after the first harmonic of the QPO. It is near 
unity at the beginning and decreases rapidly during the rising phase. 
Also observed is that the more widely separated the two energy bands 
are the less the coherence function between the two. It is interesting 
that the coherence function increases significantly at the frequencies 
of the QPO and its harmonics. We discuss the implications of the
results on the models proposed for BHCs.

\end{abstract}


\section{Introduction}
Black hole candidates (BHCs) are characterized by rapid X-ray
variability (see recent reviews by van der Klis 1995 and Cui 1999a). 
It is also common for BHCs that the variability at high energies lags
behind that at low energies (Cui 1999a and references therein), which
is often referred to as hard lag. The hard lag is often attributed 
to thermal inverse-Comptonization processes (e.g., Miyatomo et
al. 1988; Hua \& Titarchuk 1996; Kazanas et al. 1997; B\"{o}ttcher 
\& Liang 1998; Hua et al. 1999), which are generally thought to be 
responsible for producing the characteristic hard tail in the X-ray
spectra of BHCs (Tanaka \& Lewin 1995). In these models, the hard lag 
arises simply because a greater number of scatterings are required for 
seed photons to reach higher energies. Therefore, the lag is directly
related to the diffusion timescale through the Comptonizing region, 
which scales logarithmically with photon energy (e.g., Payne 1980; 
Hua \& Titarchuk 1996). The expected logarithmic energy-dependence of 
the hard lag is in rough agreement with the observations (Cui et
al. 1997; Crary et al. 1998; Nowak et al. 1999). However, the measured 
lag is often large (e.g., a few tenths of a second) at low frequencies, 
which would require a very extended Comptonizing region (Kazanas et al. 
1997; B\"{o}ttcher \& Liang 1998; Hua et al. 1999). It is not clear 
whether such a region can be physically maintained (Nowak et al. 1999; 
B\"{o}ttcher \& Liang 1999; Poutanen \& Fabian 1999). Suggestions have 
also be made to link the hard lag either to the propagation or drift 
time scale of waves or blobs of matter through an increasingly hotter 
region toward the central black hole where hard X-rays are emitted 
(Miyamoto et al. 1988; Kato 1989; B\"{o}ttcher \& Liang 1999) or to the 
evolution time scale of magnetic flares (Poutanen \& Fabian 1999). 
Regardless which scenario turns out to be close to reality, it is clear 
that the hard lag is an important property of BHCs which we can use to 
gain insight into the geometry and dynamics of accretion flows in these
systems.

Recently, however, it was discovered that a strong QPO in GRS 1915+105, 
a well-known microquasar, had a rather complex pattern of phase lag 
(Cui 1999b): while the hard lag was measured for the odd harmonics
of the signal, the even harmonics displayed {\em soft} lag. The 
pattern is puzzling because it does not fit naturally into any of the 
models suggested for BHCs. Speculation was made that the complicated 
QPO lag in this case might be caused by a change in the form of the 
wave that produced the QPO (Cui 1999b). It is, however, not clear what 
physical mechanisms could be responsible for such evolution of the wave 
form. Similar behavior was subsequently observed for some of the QPOs 
in XTE J1550-564 (Wijnands et al. 1999). Therefore, the 
phenomenon may actually be common for BHCs.

A related timing property to the phase lag is the coherence function
between two different energy bands. Only recently, however, enough 
attention is paid to the importance of this property (Vaughan \&
Nowak 1997) and efforts are made to compute it along with the phase
lag. Consequently, the results are very limited. It is nevertheless 
interesting to note that for BHCs the coherence function often
appears to be around unity over a wide frequency range --- the X-ray 
variabilities in different energy bands are almost perfectly linearly 
correlated on those timescales in Fourier domain (Vaughan \& Nowak 1997; 
Cui et al. 1997; Nowak et al. 1999). This puts additional 
constraints on the models for X-ray production mechanisms in BHCs. 
Lower coherence was observed of Cyg X-1 when the source was in the 
transitional periods between the two spectral states (Cui et
al. 1997). This could be attributed to the variation of the
Comptonizing region during those episodes on timescales less than an 
hour (Cui et al. 1997), in the context of Comptonization models (Hua 
et al. 1997). However, more data is required to verify such a
scenario.

In this Letter, we present the results from measuring the phase lag
and coherence function of X-ray variability for XTE J1550-564 during
the initial rising phase of the 1998 outburst (Cui et al. 1999, Paper
1 hereafter). In addition to the intense aperiodic variability, a
strong QPO was detected, along with its first and sometimes second 
harmonics, and the frequency of the QPO increased by almost 2 orders 
of magnitude during this period (Paper 1). We examine the timing 
properties of both the QPO and broad-band variability.

\section{Data and Analyses}
Paper 1 should be consulted for the details of the observations. Very 
briefly, there were 14 RXTE observations, covering the rising phase 
of the outburst. In the first observation, however, the overflow of 
the on-board data buffers (due to the inappropriate data modes 
adopted) produced gaps in the data. For this work, we chose to ignore 
this observation entirely. For the remaining 13 observations, we 
rebinned the data with $2^{-7}$ s time bins and combined the {\em
Event} and {\em Binned} data into the six energy bands as defined in 
Paper 1.

We chose the 2 -- 4.5 keV band as the reference band. A cross-power
spectrum (CPS) was computed for each 256-second data segment between 
the reference band and each of the higher energy bands. The results
from all segments were then properly weighed and averaged to obtain 
the CPSs for the observation. The phase of a CPS represents a phase 
shift of the light curve in a selected energy band with respect to 
that of the reference band. We followed the convention that a positive 
phase indicates that the X-ray variability in the high energy band
lags behind that in the low energy band, i.e., a hard lag. The 
uncertainty of the phase lag was estimated from the standard
deviations of the real and imaginary parts of the CPS. For the phase 
lag associated with a QPO, the magnitude was derived from fitting the
CPS in a narrow frequency range around the QPO with a linear function
(for the continuum) and two Lorentzian functions whose centroid 
frequencies and widths were fixed at those of the QPO (Paper 1). 
Acceptable fits (i.e., the reduced $\chi^2$ around unity) were 
obtained for all cases. The corresponding errors were derived by 
varying the parameters until $\Delta \chi^2 = 1$ (i.e., representing 
roughly $1\sigma$ confidence intervals; Lampton et al. 1976). 

\section{Results}
In all observations, significant hard lag was found to be associated 
with the broad-band variability. The phase lag shows significant 
evolution during the rising phase, which can be divided into two 
distinct periods. Fig.~1 shows an example for each period Fourier 
spectra of power density, phase lag, and coherence function. At the 
early stage of the rising phase, the broad-band hard lag is 
significantly measured only in the middle range of frequencies. The
lag increases with frequency first, peaks at some characteristic 
frequency, and then decreases. The peak frequency does not appear to 
correspond to any characteristic features in the power density
spectrum (PDS). At the frequency of the QPO, {\em soft} lag is clearly
detected, on top of the broad-band hard lag. However, there does not
appear to be significant phase lag (hard or soft) associated with the 
first harmonic. At the later stage of the rising phase, while the hard 
lag is still apparent there also seems to be significant {\em soft
lag} associated with the broad-band variability at frequencies around 
the QPO and its harmonics. The soft lag associated with the
fundamental component of the QPO remains significant, but now {\em
hard} lag is also measured for the first harmonic. As for the coherence 
function, it is high and nearly constant at low frequencies and drops 
off sharply almost right after the first harmonic of the QPO for both 
periods of the rising phase. It is interesting to notice that at the 
frequencies of the QPO and its harmonics the coherence function 
increases significantly. 

Both the broad-band and QPO lags show strong energy dependence, as 
illustrated in Fig.~2. They become larger at higher energies (with
respect to the reference band), which is typical of BHCs (Cui 1999a). 
The QPO lag also shows correlation with X-ray flux (and thus with the 
QPO frequencies; Paper 1) for the second period of the rising phase, 
as is apparent in Fig.~3. Both the soft and hard
QPO lags increase as the source brightens, reaching as high as 0.1 
and 0.3 radians, respectively. The coherence function is near unity
at the beginning of the rising phase. Subsequently, it decreases
rapidly and seems to level off in the end. To quantify the evolution, 
we averaged the observed values over a frequency range 0.004-0.1 Hz 
where the coherence function is roughly constant. As an example, the 
results for the 8.1-13.3 keV band are plotted in Fig.~4. The loss of 
coherence is apparent during the initial rising phase. Moreover, the 
coherence function also decreases as the separation between the two 
energy bands widens, as shown in Fig.~5.

\section{Discussion}
Although the broad-band hard lag is significantly detected only at 
intermediate frequencies, large uncertainties prevent us from
drawing any definitive conclusions on the results at low frequencies 
(where the lag is expected to be small). It appears, from Fig.~2, that 
at the early stage of the rising phase the corresponding {\em time lag} 
($t_{lag} \equiv p_{lag}/2 \pi f$) saturates below a characteristic 
frequency where the {\em phase lag} peak. This is unusual because
for other BHCs the time lag seems to monotonically increase toward 
low frequencies (e.g., Miyatomo et al. 1988; Cui et al. 1997; Grove 
et al. 1998; Nowak et al. 1999). In the context of Comptonization 
models, such a saturation in time lag might be the manifestation of 
the finiteness of the Comptonizing region (Hua et al. 1999). If so, 
the characteristic frequency (a few Hz) would provide a direct measure 
of the outer radius of the region (i.e., tens of lt-ms), which seems 
to be in rough agreement with that estimated from the measured time 
lag. It is interesting to notice that the characteristic frequency 
appears to increase as the source brightens, perhaps indicating that 
the size of the Comptonizing region decreases, as was suggested for 
Cyg X-1 when the source goes from the hard (or low) state to the soft 
(or high) state. However, it is puzzling why the lag spectrum appears 
so different at the later stage of the rising phase, or more 
specifically what causes the observed broad-band {\em soft} lag around 
the QPO and its harmonic (see Fig.~1).

The complex pattern of the phase lag associated with the QPO bear
remarkable resemblance to that observed of GRS 1915+105 (Cui
1999b). Perhaps, the phenomenon is common for certain types of 
QPOs in BHCs. Combined with the published results (Wijnands et al. 1999), 
our results show that the phenomenon seems generic for 
XTE J1550-564, as opposed to being limited to certain spectral states. 
More importantly, we are now seeing the evolution 
of the phenomenon as the QPO evolves during the rising phase of the 
outburst. If we accept the Comptonization scenario for the broad-band 
lag, we would have to rule out this scenario for the QPO because 
the QPO time lag {\em increases} with X-ray flux during the rising 
phase (see Fig.~3). This would not be surprising, since, as discussed 
by Cui (1999b), it is problematic to attribute the observed soft and 
hard lags of the QPO (and its harmonic) entirely to 
inverse-Comptonization processes in the first place. At present, no
models can naturally account for this type of phase lag phenomenon.

In the context of Comptonization models, the loss of coherence during 
the rising phase may be due to the variation in the physical
conditions of the Comptonizing region (Hua et al. 1997), as was
suggested for Cyg X-1 during spectral state transitions (Cui et al. 
1997). This scenario might also explain why more widely separated
energy bands are less coherent (see Fig.~5), since the difference in
the number of scatterings that seed photons experience is greater. 
The observed Fourier spectra of the coherence function are, however, 
somewhat unusual. For a number of BHCs, the coherence function is 
typically close to unity over the entire frequency range where it can 
be reliably determined (e.g., Vaughan \& Nowak 1997; Cui et al. 1997; 
Nowak et al. 1999). Here, the coherence function drops precipitously 
above a ``break frequency'' (which appears to be near the first harmonic 
of the QPO; see Fig.~1). We speculate that the break frequency might 
be indicative of the timescale on which the Comptonizing region varies. 
The break frequency appears to evolve in unison with the frequency of 
the QPO, which could
imply that (1) the QPO originates in the Comptonizing region (instead
of the accretion disk, as often thought) and (2) the Comptonizing
region varies on an increasingly short timescale during the initial
rising phase of the outburst. The former is also supported by the fact 
that the QPO becomes stronger at higher energies (Paper 1; also see
discussion in Cui 1999b). For BHCs in general, such energy dependence 
of QPOs is observed in most cases (Cui 1999a). Associating the QPO with
the Comptonizing region directly might also account for the intriguing
increase of the coherence function at the frequencies of the QPO and 
its harmonics, if the QPO is more localized, given that the broad-band 
variability is probably a disk phenomenon and Compton upscattering of 
disk photons in an extended non-static ``corona'' can cause the loss 
of coherence.

\acknowledgments
This work was supported in part by NASA through grants NAG5-7484 and 
NAG5-7990. We thank Markus B\"{o}ttcher for useful comments.

\clearpage
\begin{figure}
\psfig{figure=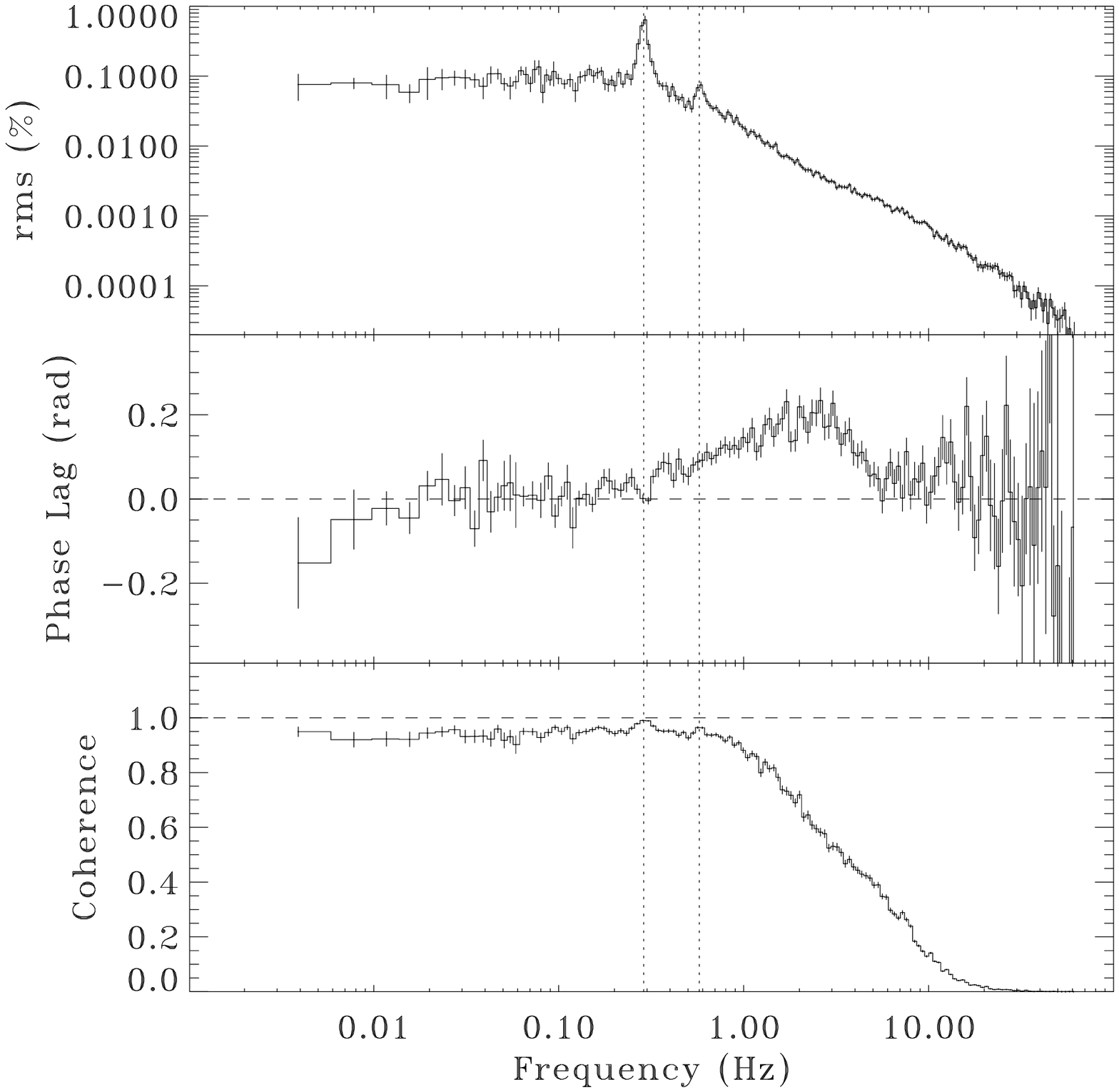,width=3.2in}
\psfig{figure=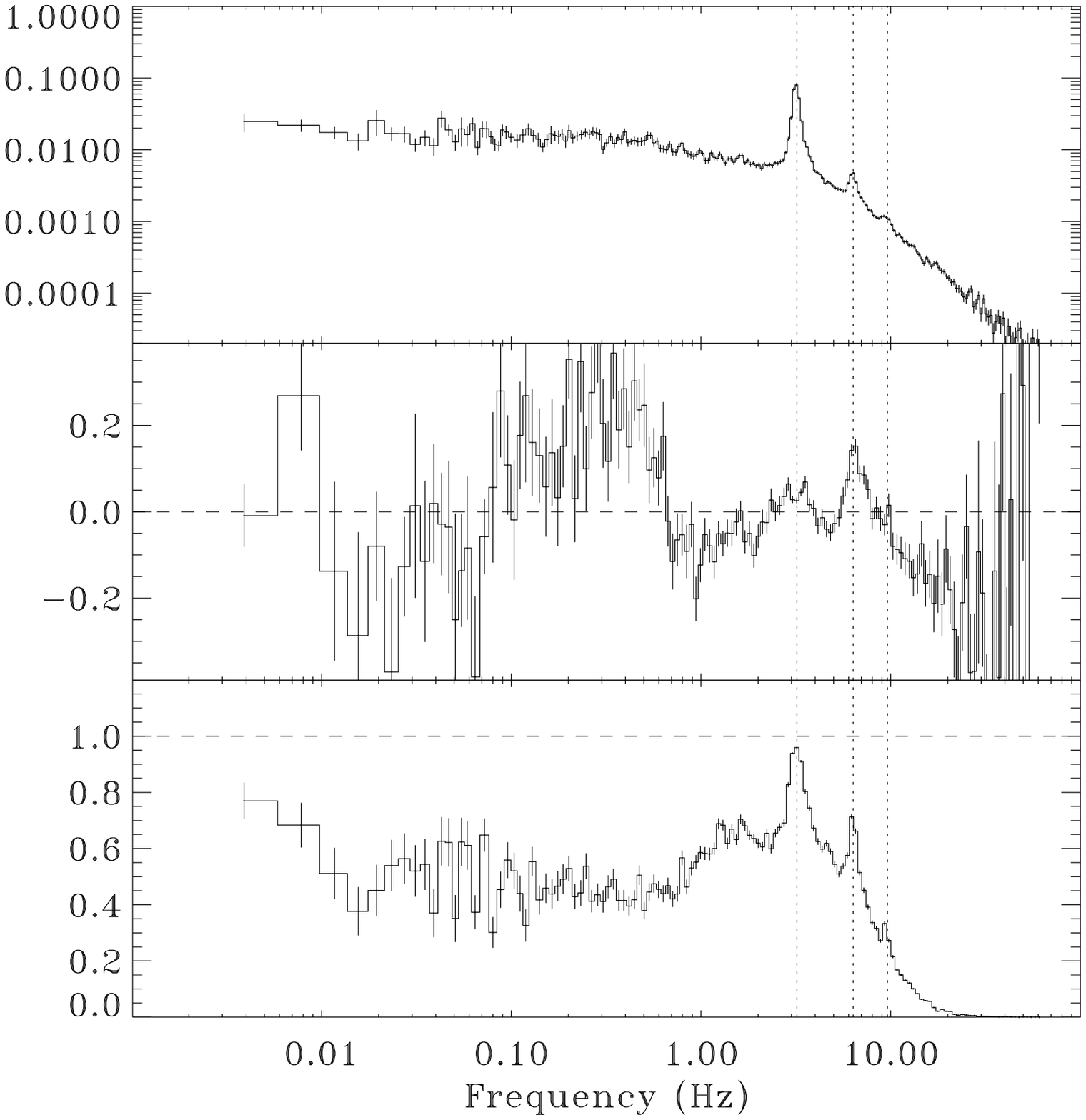,width=3.2in}
\caption{Sample Fourier spectra of power density, phase lag, and coherence 
function. The left panels show the results derived from Observation 3
(following Paper 1), and the right panels the results from Observation
10. Note the difference in the shape of the broad-band phase lag
spectrum between the two cases. The dotted lines indicate
the frequencies of the QPO and its first harmonic.}
\end{figure}

\begin{figure}
\psfig{figure=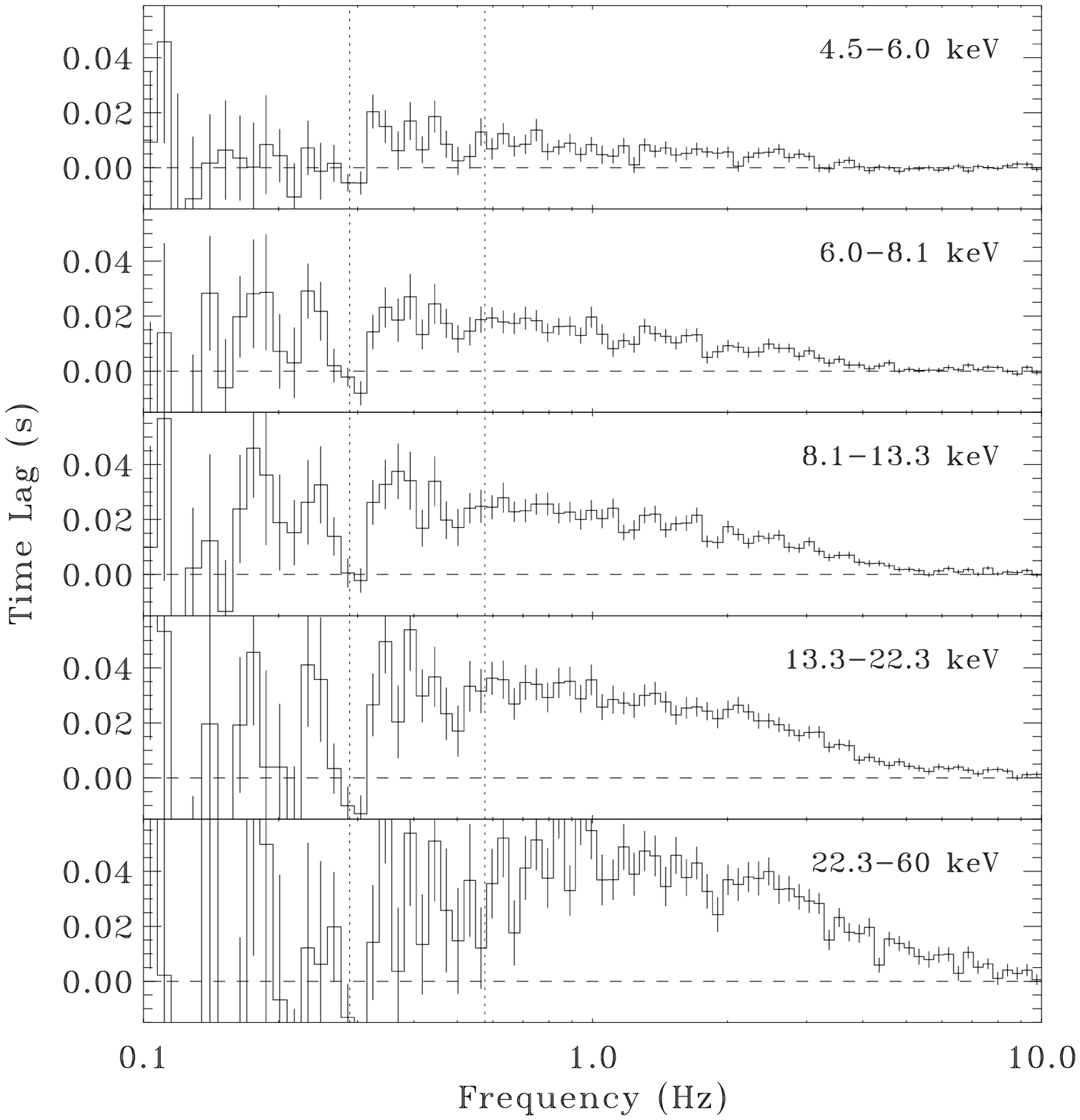,width=3.2in}
\psfig{figure=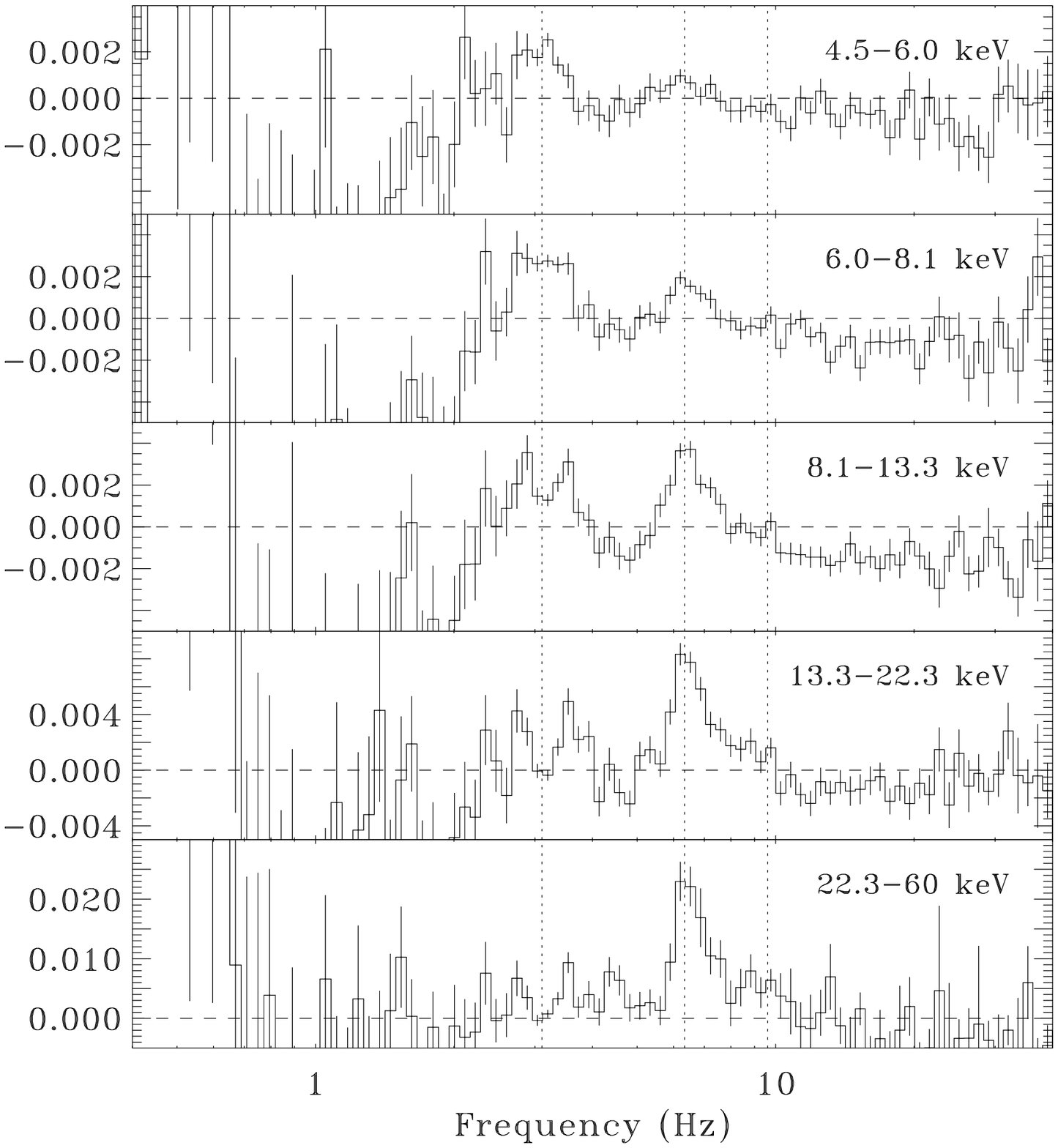,width=3.2in}
\caption{Energy dependence of time lag. As for Fig. 1, the results for 
observations 3 and 10 are shown for illustrative purposes. The dotted lines 
indicate the frequencies of the QPO and its harmonics.}
\end{figure}

\begin{figure}
\psfig{figure=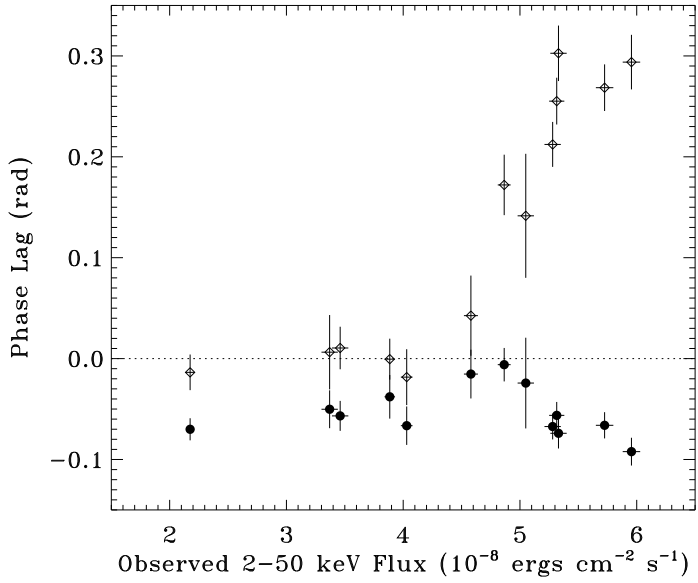,width=4.0in}
\caption{Evolution of QPO phase lag with X-ray flux. Filled circles show 
the measured ({\em soft}) phase lag associated with the fundamental 
component, while the open circles the lag with the first harmonic. 
Note that starting at the second period of the rising phase (i.e., 
Observation 8) both the soft and hard lags increase
markedly with flux.}
\end{figure}

\begin{figure}
\psfig{figure=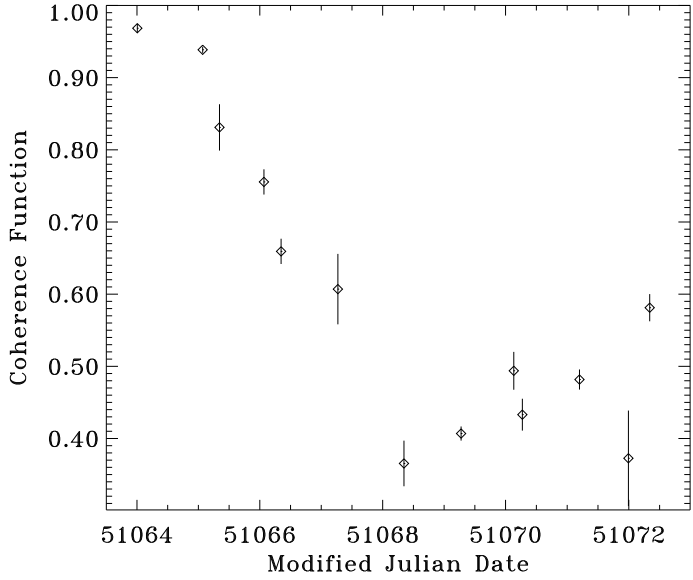,width=4.0in}
\caption{Evolution of coherence function during the 1998 outburst. The 
results shown are for the 8.1--13.3 keV band (with respect to the reference
band; see main text). They were derived by averaging over 0.004--0.1 Hz, 
where the coherence function is roughly constant. }
\end{figure}

\begin{figure}
\psfig{figure=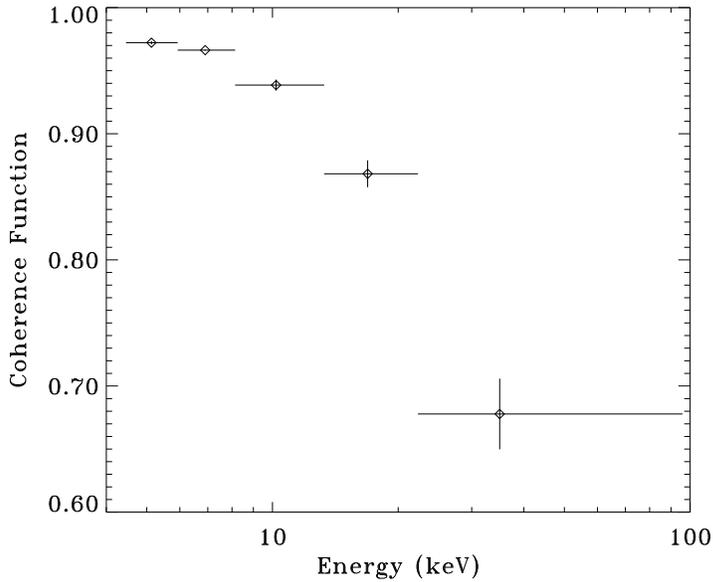,width=4.0in}
\caption{Energy dependence of coherence function. The results from
Observation 3 are shown, as an example. As for Fig. 4, the results are 
for the 8.1--13.3 keV band and derived by averaging over 0.004--0.1 Hz.}
\end{figure}
\end{document}